World Scientific
www.worldscientific.com

# Adaptive Optics for Extremely Large Telescopes

Stefan Hippler

Max-Planck-Institut für Astronomie
Heidelberg 69117, Germany
hippler@mpia.de

Adaptive Optics (AO) has become a key technology for the largest ground-based telescopes currently under, or close to beginning of, construction. AO is an indispensable component and has basically only one task, that is to operate the telescope at its maximum angular resolution, without optical degradations resulting from atmospheric seeing. Based on three decades of experience using AO usually as an add-on component, all extremely large telescopes and their instrumentation are designed for diffraction limited observations from the very beginning. This paper illuminates various approaches of the ELT, the Giant Magellan Telescope (GMT), and the Thirty-Meter Telescope (TMT), to fully integrate AO in their designs. The paper concludes with a brief look into the requirements that high-contrast imaging poses on AO.

## 1. Introduction

Over the past 30 years, ground-based astronomical observations have drawn level with their space-based counterparts. Thanks to adaptive optics (AO), the seeing limit caused by optical turbulence of the Earth atmosphere is decreasingly a real limitation for existing large optical telescopes. For the three extremely large telescopes (ELTs) currently under construction or about to begin construction, AO is an integral part of telescope design, allowing to achieve the highest angular resolution on sky: the telescope's diffraction limit.

### 1.1. *Basic principle*

The technique behind AO was first formulated by American astronomer Horace W. Babcock (1953) and can be summarized as canceling optical aberrations caused by the atmosphere in real-time.

Babcock's proposal to compensate astronomical seeing was to measure wavefront distortions (using a rotating knife-edge) and feedback that information to a wavefront correction element (Eidophor mirror). In modern applications, a fast real-time controller (RTC) sits in between a wavefront sensor (WFS) and a wavefront corrector (e.g. a deformable mirror (DM)) such that these three elements build a closed-loop system, i.e. controlling wavefronts before they reach the telescope instrument (e.g. the scientific camera). However, it took until the early 1970s before Babcock's idea could be tested experimentally (Hardy *et al.*, 1977).

### 1.2. *Early years of AO in astronomy using single natural and laser guide stars*

The usage of AO in routine astronomical observations on telescopes of the 3–4 m class can be dated back to the end of the 1980s when the instrument Come-On (Rousset *et al.*, 1990) delivered diffraction limited images in the near-infrared spectral range on a 1.5 m telescope. Upgrades from Come-On to Come-On-Plus (Gendron *et al.*, 1991) and eventually to ADONIS (Beuzit *et al.*, 1994) allowed diffraction limited imaging on a 3–4 m class telescope in the 1–5 $\mu$m spectral range.

Around 10 years later, at the end of the 1990s, the first AO system on a 8–10 m class telescope — the Keck II telescope — began operation (Wizinowich *et al.*, 2000).

First laser guide star (LGS) assisted AO systems, which were installed during the 1990s on smaller telescopes, can be seen in retrospect as prototypes, even though few scientific results have been published (Davies *et al.*, 1999; Hackenberg *et al.*, 2000). LGSs boost the operational readiness of AO-assisted observations of faint objects. An overview and historical review of sodium LGSs can be found in d'Orgeville & Fetzer (2016). LGS systems using other wavelengths than the sodium one at 589 nm, are usually called Rayleigh LGS systems, and for example the first Rayleigh system was built and used by the US military in the early 1980s (Fugate *et al.*, 1991).

On 8 m class telescopes, sufficiently bright LGS AO systems can increase the low sky coverage obtained with natural guide star (NGS) AO systems from a few percent up to about 80%, depending on the criterium for sky coverage. The most simple criterium for the sky coverage of an AO system is that the AO improves the natural seeing. This gives the highest sky coverage values. A much more stringent criterium is that the AO shall deliver fully diffraction limited point spread functions with long exposure Strehl numbers >80% at the observing wavelength. NGS AO sky coverage with sufficiently bright guide stars depends on the isoplanatic angle, while LGS AO sky coverage depends on focal and tilt anisoplanatism (van Dam *et al.*, 2006). These limitations are due to atmospheric anisoplanatism that can be partially eliminated with the usage of multiple guide stars, either natural, artificial or a combination of both.

### 1.3. Wide-field and extreme AO

A first attempt with multiple NGSs took place on the 8 m VLT on Paranal observatory in Chile in 2007. Implemented in an AO type named multi-conjugate AO (MCAO) was the MCAO demonstrator instrument MAD (Marchetti *et al.*, 2008). Results with this instrument, cf. Campbell *et al.* (2010); Meyer *et al.* (2011), showed that reconstructing and correcting wavefronts over large field of views using multiple DMs and WFSs matched the predictions. Compared to — nowadays sometimes called classical AO — single conjugate AO

(SCAO) systems with a typical non-homogeneous corrected field of view of a few 10 arcsec, MAD delivered a 2 arcmin wide homogeneously corrected field of view.

About four years later in 2011, the Gemini MCAO system GeMS, the first multiple laser and NGS AO system saw first light on Cerro Pachón in Chile (Rigaut *et al.*, 2014; Neichel *et al.*, 2014). It is currently the only operational facility instrument of its kind. GeMS has an estimated sky coverage of almost 100% in the galactic plane, >35% at medium galactic latitudes, and ∼20% at the galactic poles (Marin *et al.*, 2017), limited mainly by the required three NGSs for measuring tip and tilt aberrations over the 85″ × 85″ corrected field of view.

Another prediction of MCAO in combination with a tomographic wavefront reconstruction over the entire observing volume in the atmosphere (Ragazzoni, 1999; Ragazzoni *et al.*, 2000) was the sky coverage boost to almost 100% of the sky for an up to 100 m sized future optical telescope (Gilmozzi *et al.*, 1998) with NGSs only. Nevertheless, for the currently planned MCAO systems on the 39 m ELT and the 30 m TMT (Sec. 2), both natural and LGSs are foreseen. The expected/required sky coverage for the TMT MCAO instrument is above 70% for median seeing conditions and close to 100% under 25% best seeing conditions (Wang *et al.*, 2012). In terms of image quality, under nominal conditions, MCAO systems deliver lower Strehl ratios compared to SCAO systems.

Another type of wide field ($\geq 2' \times 2'$ up to 10–20′ × 10–20′) AO system, which delivers a corrected image image quality between seeing limited performance and MCAO performance, are so called ground-layer adaptive optics (GLAO) systems or "seeing improvers". Proposed by François Rigaut in 2002 (Rigaut, 2002), GLAO only corrects the optical turbulence located within the first ≈ 500–1000 m above the telescope pupil. The performance depends on the amount of optical turbulence within the ground layer compared to the total atmospheric optical turbulence. Using multiple NGSs, first results obtained with a GLAO system on 8 m class telescopes were achieved in 2007 (Marchetti *et al.*, 2007). Using a 5 Rayleigh LGSs GLAO system on the 6.5 m MMT telescope in Arizona (Hart *et al.*, 2010), the uncorrected seeing at 2.2 $\mu$m of 0.61″ was improved to 0.22″, about a factor of 3. Results (see Fig. 1) obtained with a 3 Rayleigh LGSs GLAO system on the 8.4 m large binocular telescope (LBT) in Arizona (Orban de Xivry *et al.*, 2016) show that a seeing improvement by a factor of 2 can be achieved under good conditions, down to 0.2″ at 1.6 $\mu$m over a 4 arcmin wide field of view.

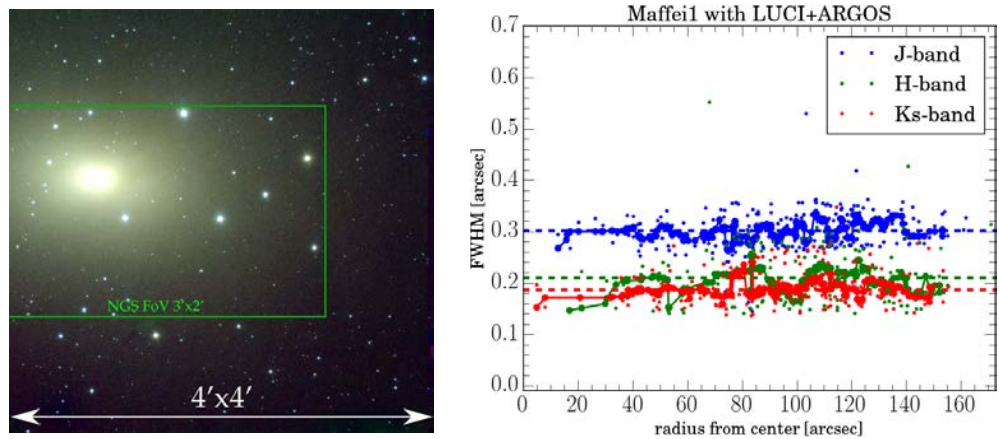

Fig. 1. Exemplary GLAO performance at the LBT with Rayleigh LGSs. (Left) LBT LUCI+ARGOS observation of the elliptical galaxy Maffei1 in J, H and K$^s$ broad band for a total of 7.5 min exposure time. The field of view (FoV) spans $4' \times 4'$ and shows a uniform resolution all over. The field-of-view available to pick a NGS for tip-tilt and truth sensing is highlighted in green. (Right) Point-spread function full width at half maximum (FWHM) as a function of radius from the center of the FoV, where was also NGS for tip-tilt sensing. The constant distribution shows that there is no apparent sign of anisoplanatism and that the GLAO correction is very uniform. Figures and slightly modified caption are taken from Orban de Xivry *et al.* (2016).

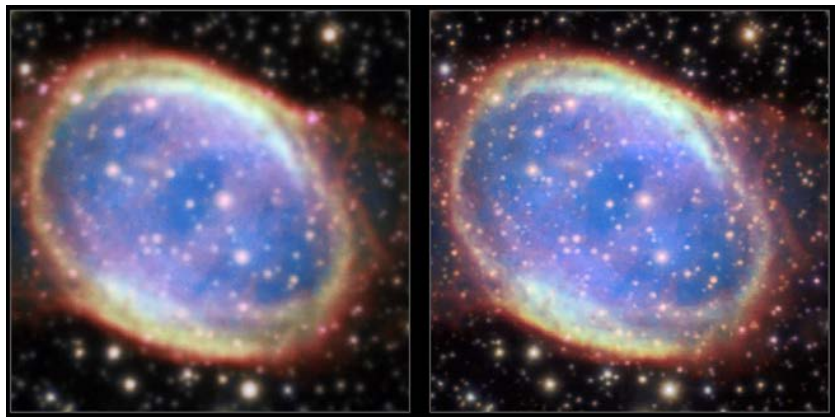

Fig. 2. Exemplary GLAO performance at the very large telescope (VLT) with sodium LGSs. VLT MUSE+AOF view of the planetary nebula NGC 6563 without GLAO correction (left) and with GLAO correction using 4 sodium LGSs (right). The field of view is $1.01 \times 1.03$ arcmin. Both images are color composites observed in the optical with various band filters ranging from 500 nm to 879 nm. (ESO, 2017).

GLAO systems using multiple sodium LGSs are currently installed and commissioned on Paranal (Paufique *et al.*, 2016; La Penna *et al.*, 2016). First results obtained with the adaptive optics facility (AOF) and the instrument MUSE — good to recognize in (Fig. 2) — show the potential of this technique in particular for wide-field GLAO spectroscopic observations. During the MUSE science verification in 2017, observations under one arcsecond seeing conditions were improved with GLAO down to ∼0.6 arcsec across the full 1 arcmin field of view (Leibundgut *et al.*, 2017).

The planning and development of second generation AO systems for 8–10 m class telescopes, often in the context of exoplanet characterization in combination with high-contrast imaging (Milli *et al.*, 2016), started at the beginning of the 21st century (Gratton *et al.*, 2004; Fusco *et al.*, 2005, 2006; Beuzit *et al.*, 2006; Feldt *et al.*, 2007; Esposito *et al.*, 2006; Macintosh *et al.*, 2006; Graham *et al.*, 2007; Martinache & Guyon, 2009; Guyon *et al.*, 2010). First science results with instruments based on these extreme AO (XAO) systems were achieved around 10 years later (Esposito *et al.*, 2010, 2013; Macintosh *et al.*, 2015; Vigan *et al.*, 2016; Maire *et al.*, 2016; Currie *et al.*, 2017). Typically, AO systems tagged "extreme" sample wavefronts at kilohertz frame rates, and correct wavefronts with a high number of actuators (for example: 672 actuators in the LBT first light AO system (Esposito *et al.*, 2006) or 4096 actuators available in the Gemini planet imager AO system (Macintosh *et al.*, 2006)).

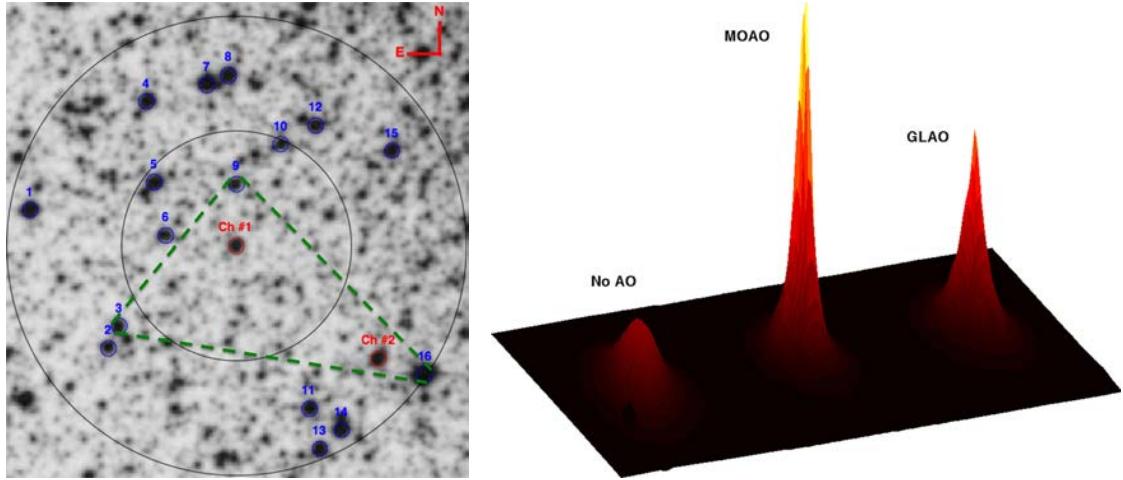

Fig. 3.   Exemplary RAVEN asterism inside the M22 globular cluster (left). The two 4 arcsec × 4 arcsec regions, science channel (Ch) #1 and (Ch) #2 lie in the green colored triangle formed by three NGSs, each feeding its own WFS. Further suitable guide stars are marked with blue circles. The diameter of the two black circles correspond to 60 and 120 arcsec. The right figure shows an exemplary PSF of a M22 star with no AO (left), and with MOAO and GLAO corrections (middle and right respectively). MOAO is shown to outperform GLAO. Both figures were taken from Lamb *et al.* (2017).

The last flavor of AO type to mention before looking at the AO-related hardware designs for the upcoming ELTs, is the multi-object adaptive optics (MOAO) system. The goal of MOAO is to compensate atmospheric turbulence over a wide field of view up to 5–10 arcmin, using individual DMs for each object to be observed. Using a small number of either natural or LGSs, similar to MCAO, the wavefront is reconstructed using tomographic methods (Ragazzoni *et al.*, 1999; Vidal *et al.*, 2010). Each DM is optimally shaped to correct the turbulence in its viewing direction, like a multi-SCAO system operated in open-loop. MOAO demonstrator systems used in the laboratory (Laag *et al.*, 2008; Ammons *et al.*, 2008) and on sky (Gendron *et al.*, 2011; Lardière *et al.*, 2014), have shown that MOAO performs close to conventional SCAO systems.

The first MOAO system installed on a 8 m class telescope was the RAVEN (Andersen *et al.*, 2012; Lardière *et al.*, 2014; Davidge *et al.*, 2015; Ono *et al.*,

2016; Lamb *et al.*, 2017) demonstrator, which was in operation at the Subaru telescope between 2014 and 2015. Developed as a pathfinder for the 30 m TMT, results obtained with this instrument are encouraging, with better performance than GLAO (Fig. 3) but not reaching SCAO performance. RAVEN can use either three NGSs or a combination of one sodium LGS and two NGSs. Two DMs with 145 actuators each, allow AO corrections on two science fields with a maximum separation of about 3.5 arcmin.

Conceptual drawings of the various AO concepts, i.e. SCAO, GLAO, MCAO, LTAO, and MOAO, including drawings of conventional WFSs and photos of DMs, can be found in the review paper of Davies & Kasper (2012).

It goes without saying that the current, often already the second generation of AO systems on 8–10 m class telescopes, has generated a wealth of experience. These experiences and lessons learned (see for example Lloyd-Hart *et al.* (2003); Wizinowich (2012); Fusco *et al.* (2014); Lozi *et al.* (2017)) are of course incorporated into the design and simulations of the AO facilities to be built. Since it would go beyond the scope of this review, here are just two current examples.

The SPHERE XAO system installed at the Paranal observatory discovered a so called low wind effect Milli *et al.* (2018) (sometimes also named island effect), which basically distorts the wavefronts in the pupil of large telescopes obstructed by spiders. The effect occurs mainly when wind speeds above the telescope are close to zero. The SPHERE XAO system cannot measure these piston aberrations, thus they propagate uncorrected to the science channel and degrade the image quality. The solutions proposed and tested will influence the design of possible upgrades to the SPHERE XAO system as well as future AO systems.

The second example concerns a general problem of most AO systems: telescope vibrations and their impact on AO performance. Experiences with the SCExAO system at the Subaru telescope (Lozi *et al.*, 2016) show weaknesses of the control system in handling vibrations, which may be mitigated by the use of predictive-control (Males & Guyon, 2018).

## 2. ELT, GMT, and TMT AO Systems

### 2.1. *AO architecture and types for first light instrumentation*

All ELTs currently under design and construction have AO kind of built-in. This means that either all components of the AO are part of the telescope or that the AO pieces are split among telescope and instruments.

The 39 m European Southern Observatory's (ESO) ELT (previously E-ELT) on Cerro Armazones in Chile has a five-mirror optical design. Its large adaptive quaternary mirror (Fig. 4) has about 5,000 actuators.

The Gregorian-type 25.4 m Giant Magellan Telescope (GMT) at Las Campanas observatory in Chile, developed by a United-States led consortium,

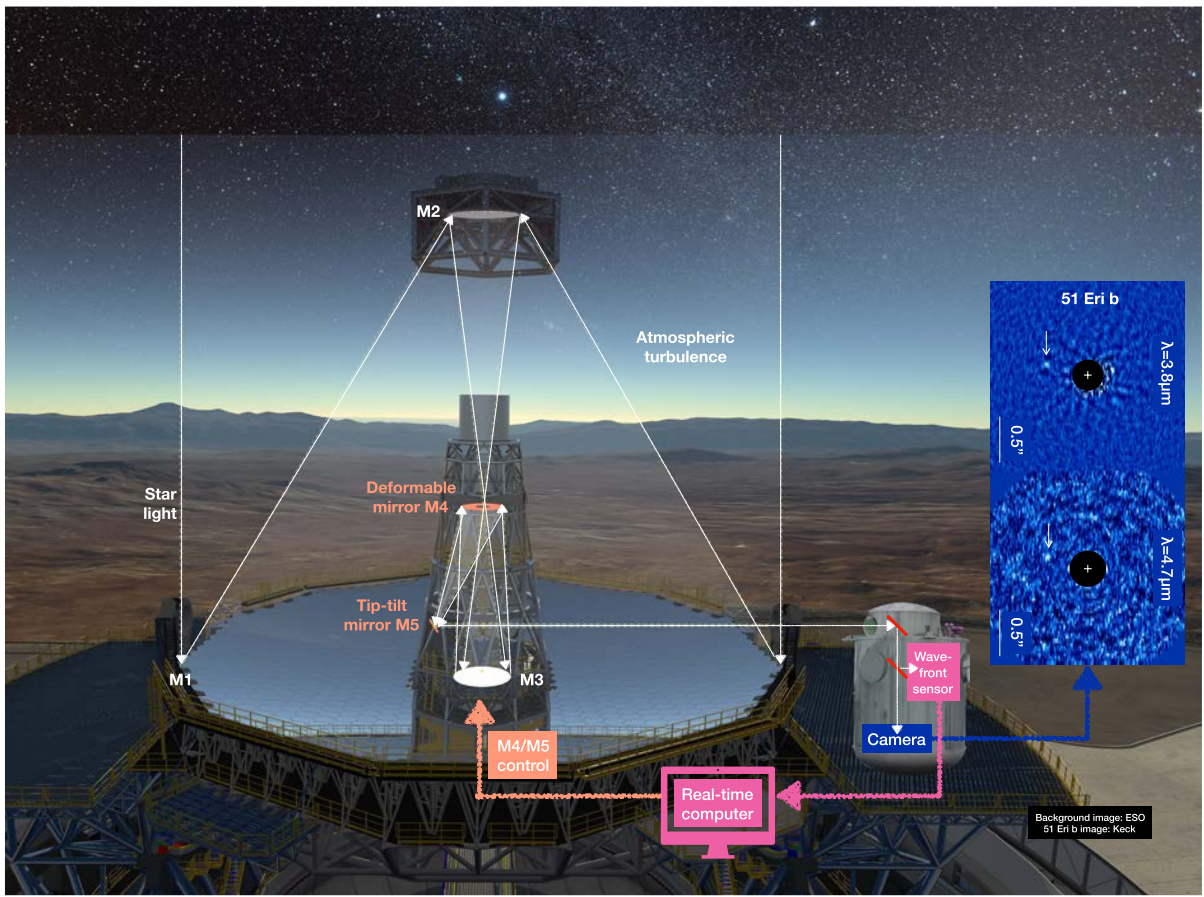

Fig. 4. Single-conjugate AO scheme for the ELT METIS instrument. Atmospheric turbulence distorts the observed starlight collected by the ELT's segmented primary mirror M1. Reflected up to mirror M2 and down to mirror M3, passing through a hole in the DM M4, the last reflection on the tip-tilt mirror M5 forwards the light into the METIS cryostat on the right, outside of the telescope. The METIS internal WFS measures the atmospheric distortions. A real-time computer estimates the actual wavefront and delivers the corresponding information to properly shape the DM M4 as well as to properly steer the tip-tilt mirror M5 with the M4/M5 control unit. Eventually, an exoplanet like 51 Eri b can be detected at two different wavelengths (see blue-colored inset on the right).

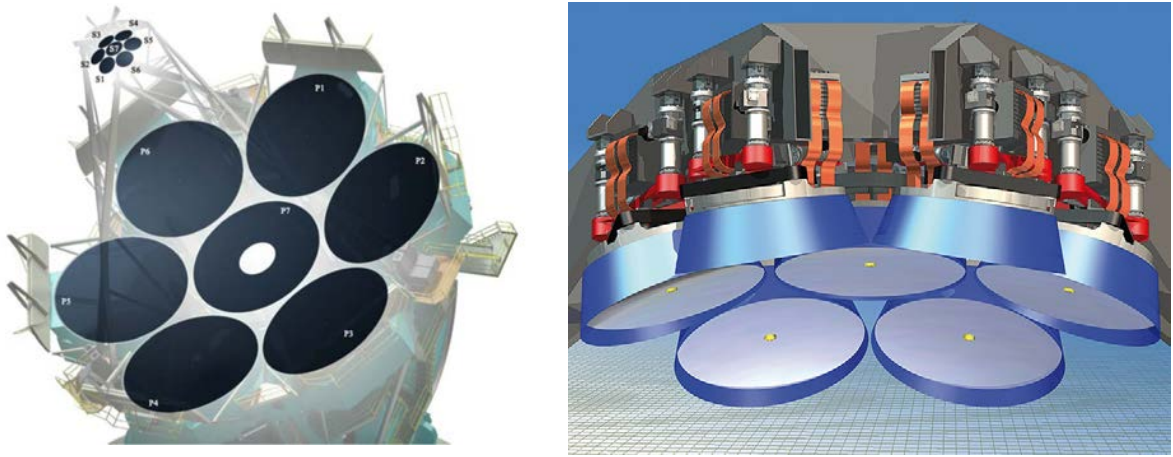

Fig. 5. GMT optical layout of the seven segment primary (P1–P7) and secondary mirror (S1–S7) on the left, a more detailed rendering of the seven segment adaptive secondary on the right. Each adaptive secondary segment has 672 contactless voice-coil actuators. Figures taken from GMTO (2013).

will have a large segmented adaptive secondary mirror (Fig. 5) with roughly the same number of actuators as the ELT.

The Thirty Meter Telescope (TMT), which will be located either on Mauna Kea, Hawaii or Canary Islands, Spain, is developed by a consortium of universities, foundations, and national observatories in the United States, Canada, China, India, and Japan. It has a Ritchey–Chrétien optical design and no adaptive mirror built-in. The TMT will probably start operation with a conventional 3 m class "active" secondary mirror and a facility AO device including DMs and WFSs. This AO facility, called NFIRAOS (see Sec. 2.2), includes DMs with about 8000 actuators in total, and will feed three instruments.

### 2.1.1. *TMT first-generation instruments*

First-generation instruments of the TMT using AO are infrared imager and spectrograph (IRIS) (Larkin *et al.*, 2016) and infrared multi-slit spectrograph (IRMS) (Mobasher *et al.*, 2010).

IRIS is a first-generation near-infrared (0.84–2.4 $\mu$m) instrument being designed to sample the diffraction limit of the TMT. IRIS will include an integral field spectrograph (R = 4000–10,000) and imaging camera (34″ × 34″). Both the spectrograph and imager will take advantage of the high spatial resolution achieved with the Narrow-Field Infrared Adaptive Optics System (NFIRAOS) at four spatial scales (0.004″, 0.009″, 0.025″, 0.05″). The features of the design will enable a vast range of science goals covering numerous astrophysical domains including: solar system science, extrasolar planet studies, star formation processes, the physics of super-massive black-holes and the composition and formation of galaxies, from our local neighborhood to high-redshift galaxies. Description for IRIS taken from TMT (2017b) and OIR-Laboratory-UCSD (2017).

The IRMS is a near diffraction-limited multi-slit near-infrared spectrometer and imager. It will be fed by NFIRAOS. IRMS will provide near-infrared imaging and multi-object spectroscopy in the spectral range of 0.97–2.45 $\mu$m. The science case for IRMS includes planetary astronomy (exoplanets), stellar astronomy (massive stars, young star clusters), active galactic nuclei (co-evolution of black holes and galaxies), inter-galactic medium (interaction with galaxies at high-z), and the high-redshift universe (population III stars, cosmic re-ionization, nature of the high-z galaxies). Description for IRMS is taken from TMT-UCR (2014).

### 2.1.2. *ELT first-generation instruments*

First-generation instruments of the ELT are HARMONI (Thatte *et al.*, 2016), MAORY (Diolaiti *et al.*, 2016), METIS (Brandl *et al.*, 2016), and MICADO (Davies *et al.*, 2016). The short descriptions of these instruments, enclosed in quotation marks below, are taken from ESO (2018).

HARMONI, the high angular resolution monolithic optical and near-infrared integral field spectrograph, "will be used to explore galaxies in the early Universe, study the constituents of the local

Table 1.   Type of AO planned for first light instrumentation. List of used abbreviations.[c–f]

| Telescope and instrument → AO type ↓ | Extremely large telescope (ELT), 39.3 m | Giant Magellan telescope (GMT), 24.5 m | Thirty meter telescope (TMT), 30 m |
|---|---|---|---|
| SCAO[a] | METIS, HARMONI, MICADO | GMTNIRS | NFIRAOS+IRIS |
| MCAO[a,b] | MICADO-MAORY | — | NFIRAOS+IRIS, NFIRAOS+IRMS |
| LTAO[b] | HARMONI | GMTIFS, GMTNIRS | — |
| GLAO[a] | — | G-CLEF, GMACS | WFOS |
| MOAO | MOSAIC[a,b] (phase A study (Morris *et al.*, 2016)) | — | TMT-AGE (feasibility study (Akiyama *et al.*, 2014)) |

*Notes:* [a]Using NGSs. [b]Using sodium LGSs.
[c]GLAO: ground-layer AO, LTAO: laser tomography AO, MCAO: multi-conjugate AO, MOAO: multi-object AO, SCAO: single-conjugate AO.
[d]HARMONI: high angular resolution monolithic optical and near-infrared integral field spectrograph, MAORY: multi-conjugate adaptive optics relay, METIS: mid-infrared ELT thermal imager and spectrograph, MICADO: multi-adaptive optics imaging camera for deep observations, MOSAIC: multi-object spectrograph for Astrophysics, inter galactic medium, and Cosmology.
[e]G-CLEF: GMT consortium large earth finder, GMTIFS: GMT integral-field spectrograph, GMACS: GMT multi-object astronomical and cosmological spectrograph, GMTNIRS: GMT near-infrared spectrograph.
[f]NFIRAOS: narrow-field infrared adaptive optics system, TMT-AGE: TMT analyzer for galaxies in the early universe, WFOS: wide-field optical spectrometer.

Universe and characterize exoplanets in great detail."

METIS, the mid-infrared imager and spectrograph, will "focus on five scientific goals: exoplanets, proto-planetary disks, Solar System bodies, active galactic nuclei, and high-redshift infrared galaxies."

MICADO, the multi-AO imaging camera for deep observations, "is the first dedicated imaging camera for the ELT. MICADO's sensitivity will be comparable to the James Webb Space Telescope, but with six times the spatial resolution."

MAORY, the multi-conjugate AO relay for the ELT, "is designed to work with the imaging camera MICADO and with a second future instrument. MAORY will use at least two DMs, including the DM of the telescope. It measures the light from a configuration of six sodium LGSs, arranged in a circle on the sky, to obtain a kind of three-dimensional mapping of the turbulence. The laser guide stars are projected from around the circumference of the telescope's primary mirror."

### 2.1.3.  *GMT first-generation instruments*

For the GMT, the suite of first-light instruments (Jacoby *et al.*, 2016) using AO consists of a fiber-fed, cross-dispersed echelle spectrograph able to deliver precision radial velocities to detect low-mass exoplanets around solar-type stars, hence the name of the instrument GMT-consortium large earth finder G-CLEF (Szentgyorgyi *et al.*, 2016). Both, the wide-field optical multi-object spectrograph GMACS (DePoy *et al.*, 2012) and G-CLEF will benefit from the GMT's ground-layer AO observing mode (Bouchez *et al.*, 2014; van Dam *et al.*, 2014). The near-infrared integral field spectrograph and imager GMTIFS (Sharp *et al.*, 2016) and the near-infrared high-resolution spectrograph GMTNIRS (Jaffe *et al.*, 2016; Jacoby *et al.*, 2016) will use the GMT's laser tomography AO system (Bouchez *et al.*, 2014; van Dam *et al.*, 2016).

An overview of all three telescopes, their AO type, and their planned first suite of instrumentation is given in Table 1.

### 2.2.  *Wavefront sensing set-up*

#### 2.2.1.  *Natural and artificial laser guide stars*

WFSs analyze light passing through the Earth's atmosphere. The ideal light source for wavefront sensing, called NGS, is a stellar point like object, as close as possible to the scientific target, and sufficiently bright to minimize measurement errors. Artificial light sources created high above the telescope — for example in the ~100 km high sodium layer of the mesosphere —, called LGSs, can be positioned as close as required to the scientific target. Their brightness depends on the used technology. LGSs enormously increase the usability of AO systems. LGSs as well as the technique of laser tomography AO are mentioned in this article but going into details is beyond the scope of this review.

Table 2. Planned LGS facilities on ELT, GMT, and TMT.

| Telescope | Laser system | Power per laser [Watt] | Comments |
|---|---|---|---|
| ELT | sodium laser CW | $\approx 22$ | At least four lasers based on the VLT four LGSF (Bonaccini Calia *et al.*, 2014). |
| GMT | sodium laser | $\approx 20$ | Six lasers for LTAO observations, D'Orgeville *et al.* (2013); Bouchez *et al.* (2014) |
| TMT | sodium laser CW or pulsed | 20–25 | Up to nine lasers, Li *et al.* (2016) |

For an overview, Table 2 summarizes for each telescope the plans for installing LGS facilities.

### 2.2.2. *Wavefront sensing architecture*

The GMT SCAO and LTAO architecture (Fig. 6) is designed to use reflected light from a tilted instrument entrance window for wavefront sensing, using either light from natural or LGSs. The reconstructed wavefront is used to shape the adaptive secondary mirror of the GMT, such delivering corrected wavefronts to the instrument. In the case of LGS wavefront sensing additional on-instrument WFSs are required to sense low-order atmospheric aberrations. For GLAO operations, the GMT has an integrated acquisition, guiding, and wavefront sensing system (AGWS) analyzing the light in an annular field of view outside the science instruments field of view. This "technical" field of view with an inner diameter of about 6 arcmin and an outer diameter of about 10 arcmin, is large enough to find suitable NGSs for GLAO operations.

In contrast to the GMT, the ELT does all AO related wavefront sensing inside the instruments. An exception from this ELT AO design is MAORY, which is designed to support two instruments. Wavefront compensation is performed using the ELT's internal adaptive mirrors M4 and M5 as shown in Fig. 4. Active optics and possible GLAO support is done in a similar way as for the GMT, i.e. using a "technical" field of view.

How does the TMT compare with the GMT or ELT AO architecture? The TMT has a facility MCAO/SCAO system called NFIRAOS (Herriot *et al.*, 2014, 2017), which can feed three instruments: an infrared imaging spectrograph IRIS, an infrared multi-slit spectrometer IRMS, and a third future instrument. From this perspective, NFIRAOS is similar to the ELT's MAORY "relay" system, which feeds MICADO and a second future instrument. In its current design, the TMT itself has no adaptive DM built-in, instead, the wavefront correction devices are located inside NFIRAOS. A laser guide star facility (LGSF) will provide at least six sodium LGSs (Li *et al.*, 2016) feeding NFIRAOS, thus enabling LGS MCAO observations (Boyer & Ellerbroek, 2016). In September 2017, a design study for an adaptive secondary mirror was launched (TMT, 2017a). In particular, the wide-field optical spectrometer (WFOS) would benefit from such an integrated wavefront correction unit, as it is not connected to the NFIRAOS unit, and the LGSF is designed to provide various asterisms, including a GLAO asterism TMT (2018a).

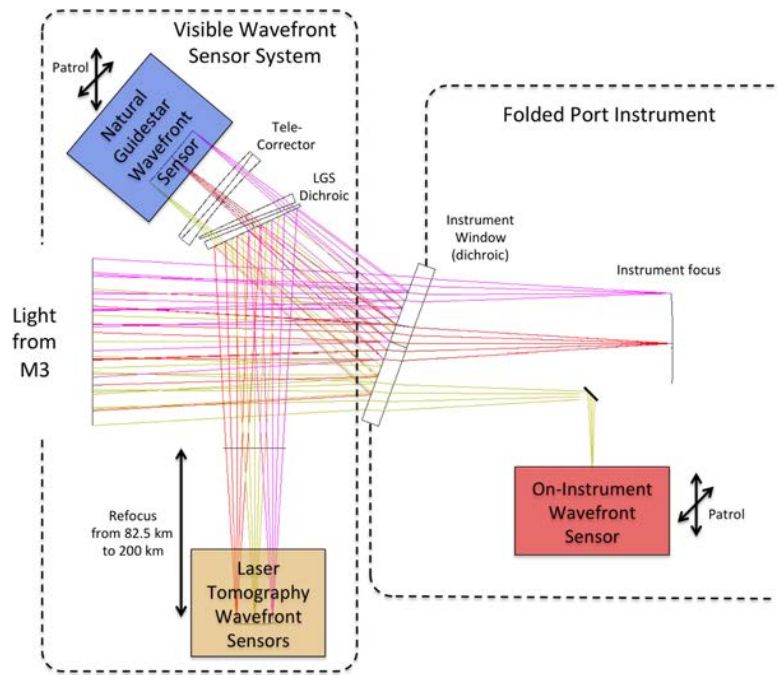

Fig. 6. The GMT LTAO, and SCAO design. Light reflected from the telescope's 3rd mirror (M3), before entering the science instrument, passes through the wavefront sensing area of the telescope. Figure taken from Bouchez *et al.* (2014).

Table 3 summarizes the main characteristics of the WFS used with the first light and first-generation respectively instrumentation.

### 2.3. *Wavefront correction devices*

This brings us to the topic of wavefront correction devices, their characteristics and location within each of the described telescopes and instruments, respectively. As shown in Fig. 4, the ELT comes with a DM M4 and a fast steering mirror M5. The $2.4 \times 2.5$ m sized elliptic mirror M4 is supported by 5316 contactless voice-coil actuators (Biasi *et al.*, 2016). This mirror can be controlled at frequencies up to 1 kHz. Together with the fast and flat steering mirror M5, which compensates for rather large amplitude atmospheric tip-tilt aberrations (field stabilization), this combination in interplay with the wavefront sensing devices, allows GLAO, SCAO, and LTAO observations. The ELT's MCAO relay MAORY adds another two DMs for instruments attached to it. Each of them with 500–1000 actuators (actuator number and technology not yet decided).

The GMT has two exchangeable secondary mirrors, each 3.25 m in diameter, segmented and shaped similar to the primary mirror (Fig. 5). The fully adaptive secondary mirror (ASM) is supported by 4704 contactless voice-coil actuators. An alternative fast steering secondary mirror (FSM) with seven rigid circular segments (Lee *et al.*, 2017) will be used during commissioning of the telescope and later on during ASM maintenance or repair.

Wavefront correction at the TMT takes place in it's NFIRAOS facility (Fig. 7) operated at a temperature of −30 degrees. A tip-tilt stage and two DMs with a total number of 7673 actuators (3125 actuators for the ground layer DM and 4548 actuators for the high-layer DM) support MCAO observations. As mentioned in Sec. 2.2, TMT organization has launched a design study for an adaptive secondary mirror in September 2017.

Table 4 summarizes some basic parameters of the planned wavefront correction elements for all three telescopes. From today's perspective, piezo actuator technology is still in use, voice-coil actuator technology has become a standard, and bimorph mirror technology has disappeared. Liquid crystal spatial light modulators do not play a role in the context of AO for the first light instrumentation of the three telescopes. The role of micro electro-mechanical systems (MEMS) based DMs is rather small, only the GMT considers a MEMS DM for some of their first light instrumentation (Copeland *et al.*, 2016).

A further increase in actuator density, i.e. for future XAO systems, from 5–10 actuators/m$^2$ to 12.5–25 actuators/m$^2$ seems possible but has to be developed (Madec, 2012; Riaud, 2012; Kasper *et al.*, 2013; Kopf *et al.*, 2017). For high-contrast XAO systems, high-density DMs might be used to remove residual speckles seen in the science focal plane. In combination with the science focal plane detectors behind coronagraphs, a second stage AO system can be set-up in order to calibrate and remove residual speckles seen in the science images (Macintosh *et al.*, 2006; Thomas *et al.*, 2015).

### 2.4. *AO real-time control systems*

The last major component of an AO system is the real-time control system (RTC), which receives wavefront measurements from the WFSs, calculates wavefront estimates, and eventually sends control signals to the wavefront correction devices like DMs and tip-tilt mirrors. For the SCAO case, the largest size of an interaction matrix using numbers from Tables 3 and 4 has about 5000 actuators and 10,000 wavefront $x$–$y$ gradients.

The time available due to AO latency requirements, i.e. the time to

- receive pixel data from the WFSs,
- pre-process the raw WFS data,

Table 3. WFSs planned for first light instrumentation of ELTs.[f] [SHS: Shack–Hartmann WFS. Pyramid: Pyramid wavefront sensor. QC: Quad-cell detector.]

| WFS → Instrument/ AO-module/ AO-mode ↓ | AO type: G = GLAO L = LTAO M = MCAO S = SCAO | Type of WFS, high-order (HO) or low-order (LO) | Wavefront sensing spectral band [$\mu$m] | Wavefront linear spatial sampling [m] | Wavefront temporal sampling [Hz] | Wavefront sensor detector |
|---|---|---|---|---|---|---|
| | | ELT | | | | |
| HARMONI[a] | S | 1 × Pyramid, HO | 0.5–1.0 | 0.5 | 1000 | CCD220[g] |
| HARMONI[b] | G, L | 6 × SHS, HO | 0.589 | 0.5 | 500–1000 | LVSM[h] |
| HARMONI[c] | L | 1 × SHS, LO | 1.4–1.8 | 19.5 | 1000 | Saphira[i] |
| METIS[a] | S | 1 × Pyramid | 1.4–2.4 | 0.5 | 1000–2000 | Saphira[i] |
| MICADO[a] | S | 1 × Pyramid, HO | 0.45–0.9 | $\lesssim$ 0.5 | 1000 | CCD220[g] |
| MAORY[a] | M | 3 × Pyramid, HO | 0.6–0.8 | $\lesssim$ 0.5 | 500 | CCD220[g] |
| MAORY[b] | M | 6 × SHS, HO | 0.589 | $\lesssim$ 0.5 | 500 | LVSM[h] |
| MAORY[c] | M | 3 × SHS, LO | 1.5–1.8 | 19.5 | 500 | Saphira[i] |
| | | GMT | | | | |
| SCAO[a,d] | S | 1 × Pyramid, HO | 0.6–0.9 | 0.28 | 1000 | CCD220[h] |
| GLAO[a,d] | G | 4 × SHS, HO | 0.5–1.0 | 0.53–1.06 | 200 | EMCCD[m] |
| LTAO[b,d] | L | 6 × SHS, HO | 0.589 | 0.42 | 500 | NGSD[h] |
| OIWFS[c,d] | S, L | 1 × QC, LO | 2.03–2.37 | 25.4 | 1000 | Saphira[i] |
| OIWFS[c,d] | S, L | 1 × SHS, HO | 1.65–1.8 | 5.08 | 10 | Saphira[i] |
| OIWFS[c,d] | S, L | 1 × SHS, HO | 1.17–1.33 | 1.59 | 0.1 | Saphira[i] |
| | | TMT | | | | |
| NFIRAOS[a,e] | S | 1 × Pyramid, HO | 0.6–1.0 | 0.31 | 800 | MIT/LL CCD[j] |
| NFIRAOS[b,e] | M | 6 × SHS, HO | 0.589 | 0.5 | 800 | PC-CCD[k] |
| OIWFS[c,e] | M | 3 × SHS, LO | 1.1–2.3 | 15 | 800 | Hawaii-1/2RG or Saphira[l] |
| ODGW[c,e] | M | 4 windows, LO | 0.84–2.4 | 30 | 800 | IRIS detector |

*Notes:* [a]Using NGSs, usually for SCAO modes or truth sensing for LGS observations.
[b]Using sodium LGSs, usually for LTAO or MCAO modes.
[c]Low-order NGS WFSs for LGS observations.
[d]The GMT wavefront sensing architecture foresees integrated WFSs (SCAO, GLAO and LTAO), and on-instrument wavefront sensors (OIWFS).
[e]The TMT wavefront sensing architecture foresees integrated WFSs (NFIRAOS: narrow field infrared AO system), on-instrument wavefront sensors (OIWFS), and instrument internal on-detector guide windows (ODGW). OIWFS and ODGW numbers are for the TMT instrument IRIS (Andersen *et al.*, 2017).
[f]ELT data taken from Correia *et al.* (2016); Neichel *et al.* (2016); Hippler *et al.* (2018); Sereni (2017); Clénet *et al.* (2016); Esposito *et al.* (2015). GMT numbers taken from Bouchez *et al.* (2014); Pinna *et al.* (2014); van Dam *et al.* (2014). TMT numbers taken from Boyer & Ellerbroek (2016); Boyer (2017); Dunn *et al.* (2016); Kerley *et al.* (2016); Herriot *et al.* (2014, 2017).
[g]Downing *et al.* (2016).
[h]Teledyne e2v large sensor visible module (LSVM) and other sensors for AO (Jorden *et al.*, 2017). For the natural guide star detector (NGSD) device see also Reyes-Moreno *et al.* (2016).
[i]Atkinson *et al.* (2016) an references therein.
[j]MIT/Lincoln labs. 256 × 256 CCD. [k]MIT/Lincoln labs. polar coordinate CCD (Adkins, 2012).
[l]TMT intention for OIWFS/ODGW (Boyer & Ellerbroek, 2016). [m]Andor or Raptor EMCCD (Bouchez, 2017).

- calculate the wavefront slopes from calibrated pixel data,
- reconstruct the wavefront using for example a matrix-vector multiplication (MVM),
- calculate the new shapes and positions of the wavefront corrections devices including tasks like vibrations control, DM saturation management, etc.,
- send control commands to the wavefront correction devices,

is rather short and of the order of 0.2–2 milliseconds (Gratadour *et al.*, 2016; Kerley *et al.*, 2016; Bouchez *et al.*, 2014). This "low latency" requirement defines the control bandwidth of the AO system as well as the stability and robustness of the closed-loop system. Looking at the matrix-vector multiplication, one of the most time consuming operations in this list, year 2016 off-the-shelf hardware is compliant with the latency requirements. Exemplary performance tests using Intel's Xeon Phi (Knights Landing, KNL) CPU with 68 cores already shows satisfactory results as shown in Fig. 8.

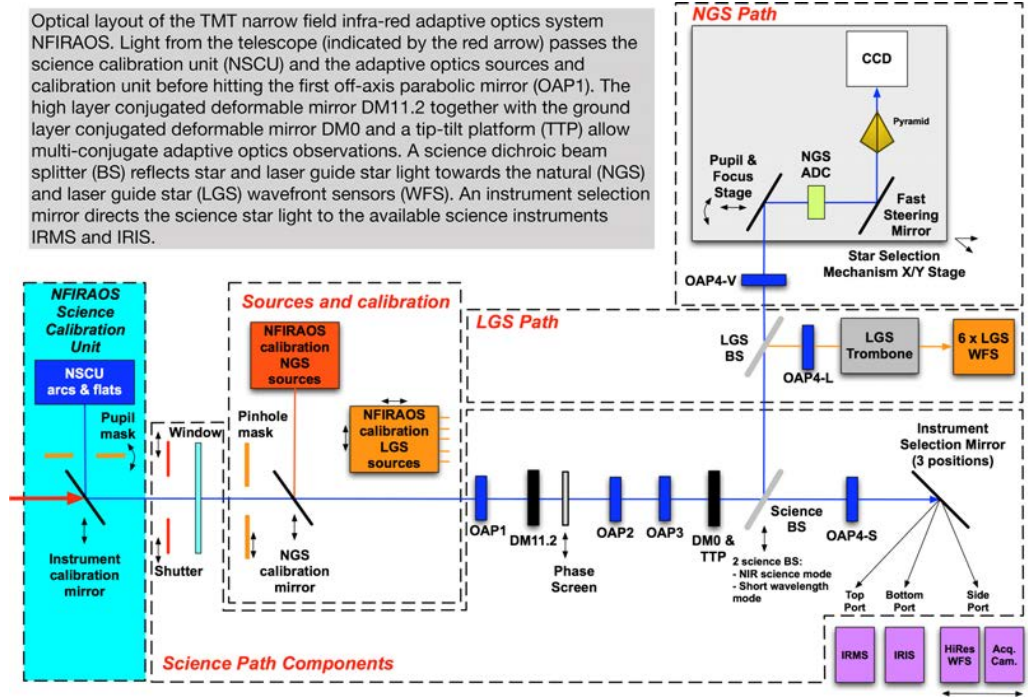

Fig. 7. Optical layout of the TMT's MCAO unit NFIRAOS. Picture without gray text box taken from (TMT, 2018a).

A schematic and simplified view of the real-time control system of the ELT instrument METIS is shown in Fig. 9. The core of METIS SCAO real-time

Table 4. Adaptive DMs planned.

| DM characteristics → Telescope location ↓ | # act., # seg.[a] | act. spacing[b] [m] | act. time[c] [ms] | act. technology M = MEMS P = piezo stack V = voice coil | Comments |
|---|---|---|---|---|---|
| GMT-M2 (ASM)[d] | 4704, 7 | 0.36 | 1 | V | Based on the LBT adaptive secondary mirror, optically conjugated to 160 m above ground. |
| GMT OIWFS-DM[e] | ≥ 20 × 20 | ≤ 1.27 | 0.5–2 | M, P, V | on-instrument DM for GMTIFS. Actuator technology not yet decided. |
| TMT-NFIRAOS-DM0[f] | 3125, 1 | ≲ 0.5 | ≤1 | P | Optically conjugated to the ground. |
| TMT-NFIRAOS-DM11[f] | 4548, 1 | ≲ 0.5 | ≤1 | P | Optically conjugated to 11.2 km height above telescope. |
| ELT-M4[g] | 5136, 6 | 0.5 | 1 | V | Optically conjugated to atmospheric ground layer. |
| ELT-MAORY[h] | ≈ 500–1000 | 1.5–2.0 | | V, P | ELT-M4 plus two DMs optically conjugated to 4 km and 14–16 km in the atmosphere. Actuator technology not yet decided (Lombini *et al.*, 2016; Pagès *et al.*, 2016). |

*Notes*: [a]Total number of actuators (act.) and segments (seg.)
[b]Actuator linear spacing over conjugated pupil [m]. [c]Actuator settling time.
[d]Biasi *et al.* (2010); Bouchez *et al.* (2014). [e]Copeland *et al.* (2016). [f]Caputa *et al.* (2014); Pagès *et al.* (2016); Sinquin *et al.* (2012).
[g]Biasi *et al.* (2016). [h]Lombini *et al.* (2016); Oberti *et al.* (2017).

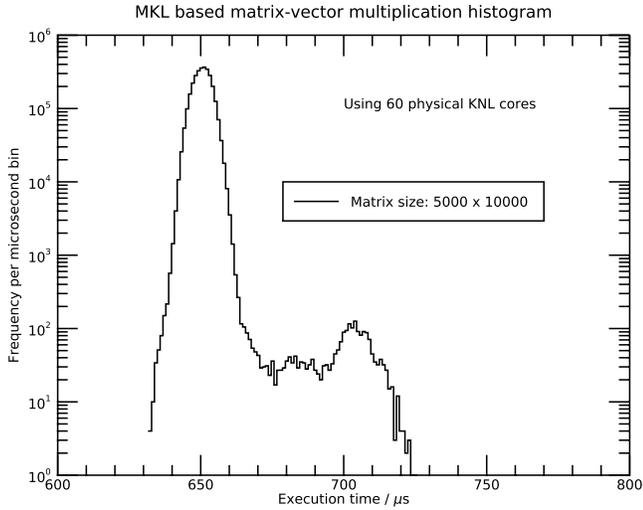

Fig. 8. Matrix-vector multiplication timing histogram using Intel's 1.4 GHz Xeon Phi 7250 (KNL) and Intel's Math Kernel Library (MKL). Matrix size: 5000 × 10000. Vector size 10000 with changing content. No. of samples: 3 million. Median execution time is ~651 microseconds with a standard deviation of ~3 microseconds.

control system (AO RTC) consists of a hard real-time controller (HRTC) and a soft real-time controller (SRTC). While the low-latency HRTC processes WFS data and generates control commands for the control systems of the correction devices, the soft real-time controller receives additional wavefront correction signals from the science focal planes (FPs) and their detector control systems (DCSs), respectively. Such additional wavefront corrections signals can for example compensate non-common path aberrations between the science and WFS light path, through adding slope offsets to the actual measured wavefront slopes. The generated correction commands can be either sent directly to the ELT's M4 and M5 local control systems (LCS) or through the ELT central control system (CCS). Further optical components within the common-path of the METIS science and WFS channel are an optical de-rotator and the pupil stabilizer (pupil stabil.). Monitoring the pupil position can be performed on the SRTC and if necessary correction commands generated and sent to the pupil stabilization controller. The field selector inside the WFS channel is pre-set via the AO observation coordination system (OCS) and the pyramid WFS modulator has to be synchronized with the SCAO WFS detector control system. Wavefront control taking into account non-common path aberrations (NCPAs) is achieved through a communication channel between the science detectors and the SRTC. The AO function control system (FCS) completes the METIS SCAO control system.

To investigate real hardware options supporting all first-generation ELT instruments, the Green Flash project (Gratadour *et al.*, 2016) aims in building a prototype/demonstrator designed to support all AO systems. After having investigated GPU clusters like the NVIDIA DGX-1 (Bernard

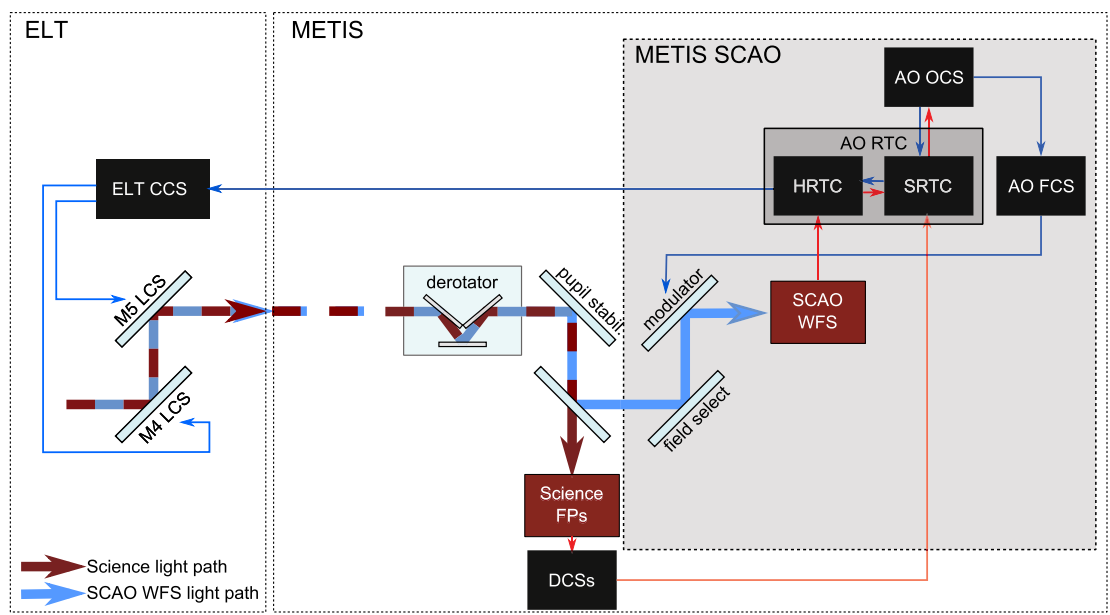

Fig. 9. Functional diagram of the METIS SCAO control system including non-common path aberrations control (Bertram *et al.*, 2018). See text for details.

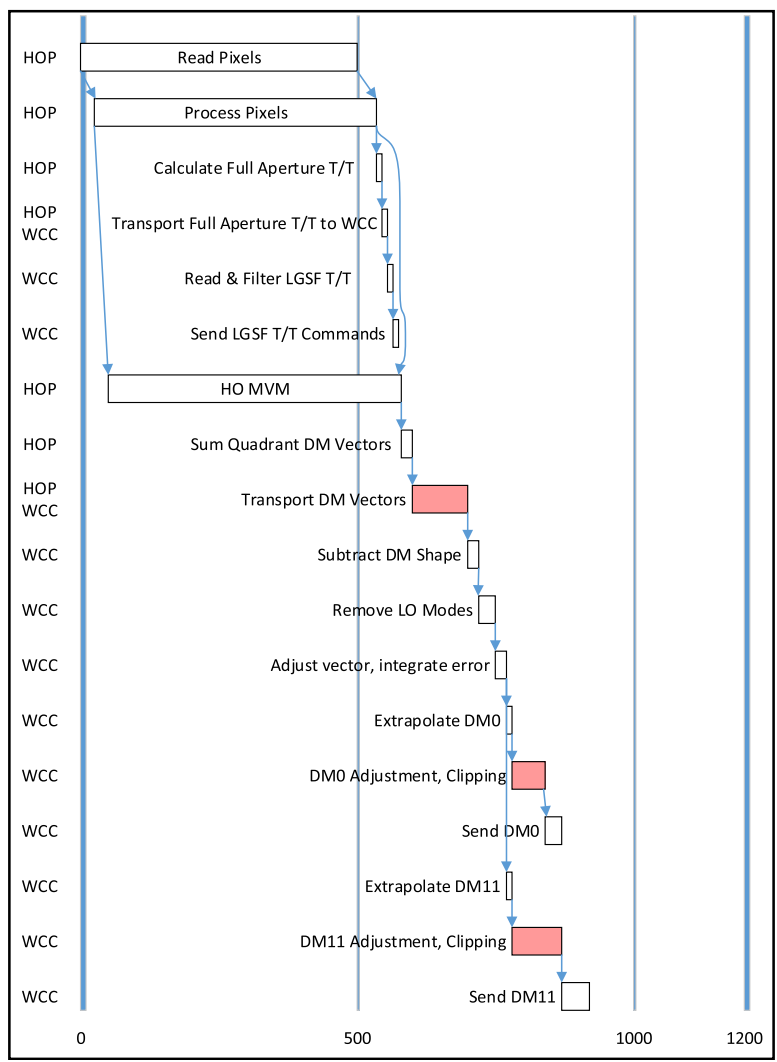

Fig. 10.   Estimated real-time pipeline stage execution times of the TMT NFIRAOS RTC. Server name (left) and task name with box indicating the execution time in milliseconds. See text for details. Figure taken from Smith *et al.* (2016).

*et al.*, 2017), Intel Xeon Phi multi-core devices (Jenkins *et al.*, 2018), and a FPGA microserver (Green Flash Project, 2018), making the choice of a single solution (if necessary) remains open.

For the TMT's MCAO system NFIRAOS, more real-time tasks, compared to the METIS SCAO RTC, are required to process the data acquired by up to six high-order WFSs. The sequence of the high-order AO data flow starts with pixel processing and wavefront reconstruction (HOP) for each WFS. A wavefront corrector control server (WCC) combines all measurements including low-order (LO) wavefront errors (full aperture NGS tip/tilt as well as LGS tip/tilt), and finally sends the correction signals to the high-order DMs DM0 and DM11 as well as to a tip-tilt stage (TTS), the LGS fast steering mirrors (FSM), and the telescope control system. The TMT RTC design allows to run the high-order pixel processing and wavefront reconstruction using a matrix-vector multiplication (MVM) almost in parallel. Estimated start and execution times for the various tasks are shown in Fig. 10.

The GMT's wavefront control system (WFCS) runs as an integral part of the telescope control

Table 5.   Exemplary AO RTC HPC hardware planned or under study. Multiple devices are necessary.

| Telescope → Hardware ↓ | ELT | TMT | GMT |
|---|---|---|---|
| CPUs | Intel Xeon Phi (KNL) | Intel E7-8870 V4 | yes |
| GPUs | NVIDIA DGX-1 | no | yes |
| FPGAs | Arria 10 | no | no |

Table 6. AO requirements (R), goals (G), and preliminary performance results (P).

| Instrument/ AO-module/ AO-mode ↓ | AO type[a] | Strehl ratio at λ | NGS mag.[b] | Corrected field of view | Contrast at distance to central PSF | Comments and references |
|---|---|---|---|---|---|---|
| **ELT** | | | | | | |
| HARMONI | S | P: 78% λ = 2.2 μm | R ≤ 13 | ≤ 6.5″ × 9″ | R (λ = 2.2 μm): 1.E-6 @ 0.2″ | Néichel et al. (2016). |
| HARMONI | L | P: ~ 40–67% λ = 2.2 μm | H ≤ 20 | ≤ 6.5″ × 9″ | | Zenith angles: 0–60°. |
| METIS | S | P: ~ 73–98% λ = 3.7 μm | K ≤ 10.8 | on-axis[c] | R (λ = 3.7 μm): 3.E-5 @ 0.1″; G (λ = 3.7 μm): 1.E-6 @ 0.1″ | Seeing conditions: 0.43–1.04″, zenith distance 30°. Hippler et al. (2018); Bertram et al. (2018). |
| MICADO | S | R: 60% λ = 2.2 μm P: 80% λ = 2.2 μm | V ≤ 12 | 50″ × 50″ | see Perrot et al. (2018) | Clénet et al. (2016), Vidal et al. (2017) |
| MAORY | M | R: ≥ 30% λ = 2.2 μm G: ≥ 50% λ = 2.2 μm | R: >50% of the sky | ≤75″ diameter | n/a | Diolaiti et al. (2017). Corrected field of view is the MICADO science field of view. |
| **GMT** | | | | | | |
| NG(S)AO | S | P: 89.3% λ = 2.18 μm | R < 8 | on-axis (NIRS) 20.4″ × 20.4″ ≤ 4.4″ × 2.25″[d] | R (λ = 3.8 μm): 1.E-5 @ 0.12″ | Zenith distance: 15°. Bouchez et al. (2014); Pinna et al. (2014) |
| LTAO | L | R: >30% λ = 1.65 μm | P: 79% of the sky | on-axis (NIRS) 20.4″ × 20.4″ ≤ 4.4″ × 2.25″[d] | n/a | Zenith distance: 15°. Sky coverage at the galactic pole. Bouchez et al. (2014). |
| **TMT** | | | | | | |
| NFIRAOS | S | P: 80.1% λ = 2.2 μm | R < 8 | on-axis | n/a | Boyer (2017) |
| NFIRAOS | M | P: 71–75% λ = 2.2 μm | R: > 50% of the sky | on-axis 34″ × 34″ (IRIS) | n/a | Sky coverage at the galactic pole. Boyer & Ellerbroek (2016); Boyer (2017) |
| NFIRAOS | M | | | 2.1′ × 2.1′ (IRMS) | n/a | NFIRAOS wide field mode with modest AO correction (Simard et al., 2012). |

*Notes:* [a]L = LTAO, M = MCAO, S = SCAO.

[b]NGS spectral band and magnitude or fraction of the sky observable.

[c]METIS requirements and estimates given for on-axis PSF. The METIS total corrected field of view is ~ 12″ × 12″ (Brandl et al., 2018).

[d]GMTIFS field of view. The larger field of view of 20.4″ × 20.4″ is for the IFS imager (Sharp et al., 2016).

system Bouchez *et al.* (2014). The most demanding real-time requirements are set by the LTAO mode with a latency $\leq 200\,\mu s$. The GMT's baseline WFCS design foresees nine commodity servers, one slope processor for each of the six laser tomography WFSs, one node for communication with the adaptive secondary mirror, one node for communication with the on-instrument DM, and one master node for combining all reconstructed wavefronts.

Whether commercial off-the-shelf components and commodity servers can be used for the required high performance AO controllers is currently under investigation. Table 5 lists hardware components under consideration at the particular telescopes. The table brings to focus the high-performance computing (HPC) power only, while other hardware components like low-latency, high bandwidth networks (e.g. 10-100 Gbit-Ethernet, Infiniband, Omni-Path), sufficiently fast input/output channels (e.g. Camera Link, GigE Vision, sFPDP, USB3), and high throughput data storage are considered uncritical as it can be categorized as standard hardware.

## 2.5. *AO performance: Requirements and expectations*

Usually the scientific program determines the necessary minimum requirements for the measuring instrument. As an example, for the ELT METIS instrument various dedicated AO related top-level requirements influence the AO design. These are among others requirements on Strehl and contrast, for example, METIS shall deliver a minimum Strehl ratio of 60% (goal 80%) in L-band ($\lambda = 3.7\,\mu m$). For the contrast, METIS shall deliver a $5\sigma$ contrast in L-band of $3\times10^{-5}$ (goal: $1\times10^{-6}$) at a distance to the central PSF of $5\lambda/D$ (goal: $2\lambda/D$).

Detailed end-to-end simulations are in the design phase the only way to figure out whether the design meets the requirements. Table 6 lists requirements on the AO for the instruments described in Table 1 and, where available, simulated performance numbers.

## 3. High-Contrast Imaging Requirements on Adaptive Optics

High-contrast imaging astronomy aims at uncovering faint structures and substellar objects very close to bright stars (an overview on this topic gives for example Oppenheimer & Hinkley (2009)). Investigating distances close to the diffraction limit, i.e. $\sim$10–1000 mas, requires sophisticated AO installations, like XAO systems, that are able to spatially separate the light from, for instance, a host star and an orbiting planet.

A good measure to characterize the quality of an AO system is to look at the long and short exposure Strehl ratios it delivers. Together with the residual image motion, also for long and short exposures, one can use these static and dynamic performance data to estimate the imaging quality and the raw contrast delivered to the science channel. Per definition, the Strehl ratio defines the ratio of the measured AO corrected peak intensity of the point spread function (PSF) to the intensity measured or estimated of a perfectly imaged point spread function. If the delivered long exposure Strehl ratio at a wavelength of $3.7\,\mu m$ is 95%, there is a fraction of 5% of intensity or energy in the halo of the stellar PSF. As the structure and the dynamic behavior of this un-controlled halo is unknown, it contributes strongly to the achievable contrast. This becomes clear if we look at the standard technique used in high-contrast imaging, where a stellar reference PSF is subtracted from the actual recorded stellar PSF.

In combination with a coronagraph, contrast values of point sources below $10^{-4}$ at angular separations of $5\lambda/D$ on a $D = 8$ m telescope can be achieved. As shown in Fig. 11, using the AO instrument NACO at the 8 m VLT with an annular groove phase mask (AGPM), different contrast values can be achieved depending on the applied data reduction algorithm. The pure AO intensity profile of a standard star, which shows structures around the 3rd bright Airy ring and at the edge of the AO control radius, can be reduced by a factor of approx. 20 at an angular separation of 0.5 arcsec ($5\lambda/D$) with $\lambda = 3.8\,\mu m$. The subtraction of the stellar PSF using the angular differential imaging (ADI, Marois *et al.* (2006)) method together with a principal components analysis (PCA, Gomez Gonzalez *et al.* (2016)) results in higher contrast values towards smaller angular separations. The rather flat contrast values at angular separations beyond the AO control radius indicate the highest contrast achievable at the smallest angular separations for this observation set-up. As the green curve in Fig. 11 shows, there are still about 2 orders of magnitude contrast improvements possible.

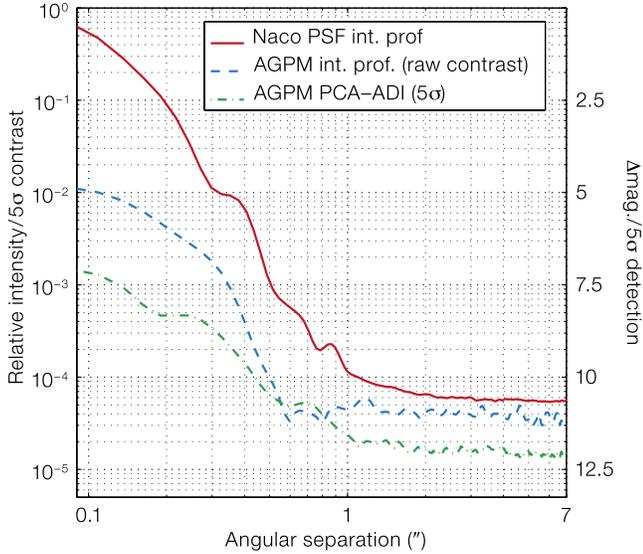

**Fig. 11.** VLT NACO coronagraphic observations of HD 123888 in L'-band ($\lambda = 3.8\,\mu m$). Normalized azimuthally averaged relative intensity profiles and contrast curve on a log–log scale. The plain red curve shows the intensity profile of a typical saturated NACO L' PSF (similar brightness and exposure time). The blue dashed curve shows the AGPM intensity profile before PCA, demonstrating the instantaneous contrast gain provided by the coronagraph at all spatial frequencies within the AO control radius ($\sim$0.7 arcseconds). The green dash-dot curve presents the reduced PCA-ADI $5\sigma$ detectability limits (40 frames, 800 s, position angle difference $\approx 30°$), taking both the coronagraph off-axis transmission and the PCA-ADI flux losses into account. Figures and caption taken from Mawet *et al.* (2013). See text for further explanations.

For that reason, it is important to understand in detail the contributions of the non-corrected aberrations to this regime.

As Mawet *et al.* (2012) point out, in particular low-order aberrations contribute to the contrast at angular separations between 1 and $4\,\lambda/D$. One proposal to further improve the raw contrast behind an AO fed coronagraph is predictive control. Males & Guyon (2018) show that the raw contrast for bright stars can be increased by more than three orders of magnitude at an angular separation from the PSF center of $1\,\lambda/D$, for $\lambda = 800\,nm$ and $D = 25.4\,m$ (GMT case). This corresponds to a raw contrast improvement of $\sim$50 at $\lambda = 3.8\,\mu m$.

Another way to further reduce low-order aberrations is to increase the speed of the AO system in case there are enough photons available from the AO reference source. An exemplary case for the bright (K = 4.5 mag) star 51 Eridani is shown in Fig. 12. Increasing the AO loop frequency from 1 to 2 kHz improves the coronagraphic contrast by a factor $\approx 3$.

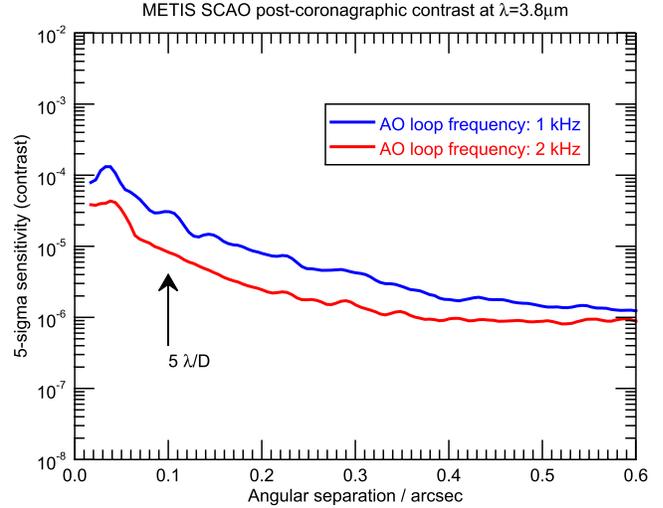

**Fig. 12.** ELT METIS 10 s simulated SCAO coronagraphic observations of the bright reference star 51 Eridani at $\lambda = 3.8\,\mu m$ (Absil & Carlomagno, 2018). This plot is to demonstrate that rather small changes of the long-exposure Strehl number from 0.978 at 1 kHz AO loop speed to 0.984 at 2 kHz AO loop speed results in a contrast enhancement of a factor $\sim$3 at separation of $5\lambda/D$. For comparison, in Fig. 11 this separation is at $\sim$0.5 arcsec.

Guyon *et al.* (2012) have shown similar behavior for a perfect AO system on a 30 m class telescope, simulating wavefront sampling frequencies up to 100 kHz and limiting the closed-loop servo latency to 0.1 ms. As shown in Fig. 13, raw H-band PSF contrasts at small separations between 10 to 20 mas of $\approx$ 1.e-5 can be achieved.

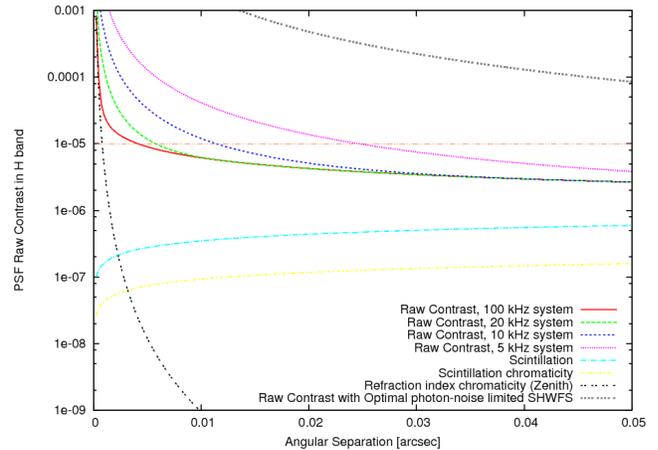

**Fig. 13.** PSF raw contrast of a 30 m telescope at $\lambda = 1.6\,\mu m$ (H-band) vs. angular separation for wavefront sampling frequencies between 5 and 100 kHz. Wavefront sensing at $\lambda = 0.8\,\mu m$ (I-band) using an ideal WFS. Shown are individual contributions to the contrast due to atmospheric scintillation, scintillation chromaticity, and atmospheric refraction chromaticity at zenith. A Shack–Hartmann wavefront sensor (SHWFS) not taking advantage of the 30 m telescope diffraction limit cannot deliver the performance of an ideal WFS. See text for further explanations. Figure taken from Guyon *et al.* (2012).

Table 7.   Planned high-contrast instruments on ELT, GMT, and TMT.

| Telescope | Instrument name | Year | Comments |
|---|---|---|---|
| ELT | MICADO, near-infrared wide-field imager | $\gtrsim 2024$ | 1st generation instr., Perrot *et al.* (2018). |
| ELT | METIS, mid-infrared ELT thermal imager and spectrograph | $\gtrsim 2025$ | 1st generation instr., Kenworthy *et al.* (2016). |
| ELT | Planetary camera and spectrograph (PCS, also called EPICS) | $\gtrsim 2028$ | 2nd generation instr., Kasper *et al.* (2013). |
| GMT | GMagAO-X, visible and near-infrared exoplanet imager | $\gtrsim 2025$ | 1st generation instr., Close *et al.* (2017); Bernstein (2018) |
| TMT | b/blue Michi, a mid-infrared imager and spectrometer | $\gtrsim 2030$ | 2nd generation instr., Packham *et al.* (2012); Packham (2017). |
| TMT | Planetary Systems Instrument (PSI) | $\gtrsim 2030$ | 2nd generation instr., successor of the Planet Formation Imager (PFI) concept. TMT (2018b); Dumas (2018). |

This indicates that the stability of the AO controlled PSF is a critical factor for high-contrast imaging using coronagraphs and further post-processing techniques. This is long known but the impact on contrast is striking. In particular PSF subtraction in combination with ADI requires careful execution as it can subtract extended structures like circumstellar disks Soummer *et al.* (2012). Measuring the reference PSF for post-processing PSF subtraction is a critical task in this context and requires the AO to perform as uniform as possible.

Table 7 provides an overview of the planned high-contrast instruments and the expected years of commissioning. Some of these instruments will also use the technology of high dispersion coronagraphy (HDC) not further explained here. A recent study on how exoplanets can be observed with HDC from the ground as well as with space telescopes can be found in Wang *et al.* (2017).

## 4.   Conclusions and Outlook

AO technology has evolved enormously over the last 30 years. Designed as an anytime usable option for all ELTs currently in planning and construction, it will allow diffraction-limited observations in the near infrared range and at longer wavelengths. Expanding the use of AO in the visible spectrum remains challenging — a niche observing mode for the very bests nights as Close *et al.* (2017) suggest —, but we should keep our eyes on this.

All instruments under design for the ELT, GMT, and TMT can use AO, hence make use of the huge light collecting power and the maximum achievable angular resolution. LGS facilities push the AO sky coverage to levels well above 50%.

Highly specialized instruments will further boost ground-based diffraction limited high-contrast imaging and characterization of exoplanets to unprecedented planet-star contrast ratios. The entire nearby Alpha Centauri system, including the planet Proxima Centauri b (Kreidberg & Loeb, 2016) orbiting in the habitable zone, has received strong attraction in the science community, accompanied by an intense public attention. The pure imagination to find nearby Earth-like planets in habitable zones has triggered ideas to fly there for further investigation (Popkin, 2017).

## Acknowledgments

The author would like to thank his colleagues Markus Feldt, Dietrich Lemke, and Kalyan Radhakrishnan for their advice and comments. Special thanks to the anonymous reviewer for his/her very valuable comments and suggestions to improve the quality of this review.

The author also thanks the following organizations and journals for allowing him to reproduce the figures listed below in this publication. The International Society for Optics and Photonics (SPIE) for Figs. 1, 6 and 9, the Monthly Notices of the Royal Astronomical Society for Fig. 3, the ESO for Fig. 2 and the background of Fig. 4, the Giant Magellan Telescope Organization (GMTO) for Fig. 5, the Thirty Meter Telescope International Observatory (TIO) for Fig. 7, and The Messenger for Fig. 11.